\pdfoutput=1 
\documentclass[fleqn,usenatbib]{mnras}

\usepackage{newtxtext,newtxmath}

\usepackage[T1]{fontenc}
\usepackage{ae,aecompl}

\usepackage{graphicx}	
\usepackage{amsmath}	
\usepackage{amssymb}	

\title[CO in the C1 globule of the Helix nebula]{CO in the C1 globule of the Helix nebula with ALMA}

\author[Andriantsaralaza et al.]{
M. Andriantsaralaza,$^{1}$\thanks{E-mail: miora.andriantsaralaza@gmail.com}
A. Zijlstra$^{1,2}$\thanks{E-mail: a.zijlstra@manchester.ac.uk}
and A. Avison$^{1,3}$
\\
$^{1}$Jodrell Bank Centre for Astrophysics, The University of Manchester, Manchester M13 9PL, UK\\
$^{2}$Department of Physics \& Laboratory for Space Research, University of Hong Kong, Hong Kong\\
$^{3}$UK Atacama Large Millimeter/submillimeter Array Regional Centre Node, Manchester M13 9PL, UK\\
}

\date{Accepted XXX. Received YYY; in original form ZZZ}

\pubyear{2019}

\begin{document}
\label{firstpage}
\pagerange{\pageref{firstpage}--\pageref{lastpage}}
\maketitle

\begin{abstract}
We present and analyse $^{12}$CO, $^{13}$CO and C$^{18}$O(2--1) ALMA observations of the C1 globule inside the Helix nebula in order to determine its physical properties. Our findings confirm the molecular nature of the globule with a multi-peak structure. The $^{12}$CO line has a high optical depth $\tau\sim$10. The derived $^{12}$C/$^{13}$C$\sim$10 and $^{16}$O/$^{18}$O$\sim$115 ratios are not in agreement with the expected isotopic ratios of carbon-rich AGB stars. Assuming that the $^{12}$CO optical depth has been underestimated, we can find a consistent fit for an initial mass of 2\,M$_\odot$. We obtain a molecular mass of $\sim$2$\,\times 10^{-4}\,$\(\textup{M}_\odot\) for the C1 globule, which is much higher than its mass in the literature. Clumping could play a role in the high molecular mass of the knot. The origin of the tail is discussed. Our findings show that the most probable model appears to be shadowing. The kinematics and molecular morphology of the knot are not consistent with a wind-swept model and the photoevaporation model alone is not enough to explain the nature of the globule. We propose an integrated model where the effects of the photoevaporation, the stream and shadowing models are all considered in the tail shaping process. 

\end{abstract}



\section{Introduction}

The Helix nebula (NGC 7293) is one of the closest known planetary nebulae (PN) at a distance of 201$\pm 4$ pc \citep{gaia}. One of the most interesting characteristics of the Helix nebula is the presence of microstructures known as cometary knots inside its molecular envelope (e.g. \citeauthor{huggins1996} \citeyear{huggins1996}). Those knots, first reported by \citet{1968IAUS...34..256V}, are dense clumps of gas and dust. They exhibit a crescent-shaped head in optical images (e.g. \citeauthor{o2002knots} \citeyear{o2002knots}, \citeauthor{meaburn1992} \citeyear{meaburn1992}) and may possess a tail pointing away from the central star (CS). In this paper, the terms cometary knots, knots and globules are interchangeable and do not distinguish whether the knots present a defined tail or not.  
\\Globules are thought to be common structures inside PNe, especially in those with a molecular envelope \citep{huggins1996} such as the Ring and the Dumbbell nebulae \citep{o2002knots}. With a typical size of around $10^{13}$m (e.g. \citeauthor{nishiyama2018} \citeyear{nishiyama2018}), the knots in the Helix nebula can be resolved and studied in detail due to its proximity. \citet{matsuura2009} estimated the number of globules in the  Helix nebula to be around $40\,000$, with different distributions for globules located closer to the CS, which are more or less isolated, and those far from the CS, which are more overlapping in images. 
\\
\\Understanding the nature and behaviour of the knots is of great importance to understand the physical processes reigning inside the planetary nebula host. Over the past years, optical analyses have permitted the estimation of properties such as the kinematics, the morphology and the lifespan of the knots (e.g. \citeauthor{meaburn1998} \citeyear{meaburn1998}, \citeyear{meaburn1996}, \citeyear{meaburn1992}).  Studies have also been conducted in H$_2$ (e.g. \citeauthor{matsuura2009} \citeyear{matsuura2009}) and CO (\citeauthor{huggins1992} \citeyear{huggins1992}, \citeauthor{young1997} \citeyear{young1997}) to derive the molecular characteristics, such as the mass, the number and the density of systems of knots as well as of individual globules such as the C1 globule \citep{huggins2002}. One of the biggest uncertainties on the knots is their formation and evolution. Though a number of models have been suggested to explain how they formed (e.g. \citeauthor{dyson2006} \citeyear{dyson2006}, \citeauthor{lopez2001} \citeyear{lopez2001}, \citeauthor{canto1998} \citeyear{canto1998}), whether they have been created during or after the AGB phase and how they obtained their comet-like shape remains unclear. 
\\
\\The aim of this paper is to determine the physical properties of the globules in the Helix nebula by investigating the CO emission from an individual knot: the C1 globule.  C1 is a near-side globule located at $+\,136^{\prime\prime}$ North to the CS \citep{huggins1992}. We report and analyse new CO data of the knot with significantly better resolution than previous observations. 
\\This paper is organized as follows. In Section \ref{sec:data}, we present the ALMA data used in this work and describe the data reduction process. Our results are presented in Section \ref{sec:result}, in which the line intensities, the morphology and the optical depth are discussed. The following sections discuss the isotopic ratios, the molecular mass, the dust emission and the origin of the globules. Section \ref{sec:Conclusion} closes the paper with a conclusion. 
\section{Data and data reduction}
\label{sec:data}
The data used in this analysis were obtained from the ALMA archive under the project code 2012.1.00116.S. The C1 globule was observed with 40 12-m ALMA dishes at band 6 between 2014 July 21 and 27 (PI: Huggins), with a maximum baseline of 820.2m. The observations covered a field of view of $\sim$28 arcsec with an angular resolution of $0.39\,$arcsec and a total integration time of $2.9\,$h. The C1 globule in the Helix nebula is located at a right ascension of $22$h $29$m $37.833$s and a declination of $-20^{\circ}$ $47^\prime$ $55.017^{\prime\prime}$ (J2000).\\
\\We manually calibrated the data using \textsc{casa} \citep{mcmullin2007casa}. The retrieved data consist of four different measurement sets which we calibrated independently. 
\\For each measurement set, we initially performed calibration of the system temperature of the antennas and atmospheric water vapour. The complete measurement sets have twenty-four spectral windows from which were extracted the spectral windows centred on $230.573$, $220.432$ and $219.596$ GHz, corresponding to $^{12}$CO(2--1), $^{13}$CO(2--1) and C$^{18}$O(2--1), respectively, and on $231.635\,$GHz for the continuum emission. Each spectral line window has a width of $117.187\,$MHz over $3\,840$ channels with a velocity resolution of $0.05\,$km$\,$s$^{-1}$, whereas the continuum window covers a 1.875$\,$GHz width with 128 channels.
\\The data were flagged to exclude shadowed antennas and end channels with poor receiver response. The data were then calibrated for bandpass, flux scaling, amplitude and phase of the complex visibilities using standard CASA procedures. The calibrators used were the quasar J2258-279 and the Seyfert galaxy J2258-2758 which were used as flux and bandpass/phase calibrators, respectively. 
\\The calibrated data from the four individual sets were then combined into one calibrated measurement set. 
\\
\\After calibration, spectral cubes were extracted for each molecular line across all channels using the multiscale deconvolution algorithm, along with a continuum map obtained with the H\"{o}gbom algorithm of the \textsc{casa} imaging task \textit{tclean}. The chosen size of a pixel is $0.07\,\textmd{arcsec}$ and Briggs weighting was applied to all images with $robust\,$=$\,0.5$. The resultant data cubes have beam sizes of $0.41\times0.35$, $0.42\times0.36$ and $0.43\times0.36$ $\textmd{arcsec}\,\times\,\textmd{arcsec}$, and typical rms noises ($\sigma_{12}$, $\sigma_{13}$, $\sigma_{18}$) of $3.5$, $2.6$ and $2\,\textmd{mJy}\,\textmd{beam}^{-1}$ for $^{12}$CO, $^{13}$CO and C$^{18}$O, respectively. The size of the synthesised beam for the continuum map is $0.40\times0.34$ in arcsec, with an rms noise of $\sim20\,\mu\textmd{Jy}\,\textmd{beam}^{-1}$. Continuum subtraction was not applied when producing the line cubes as the continuum emission is weak. For creating ratio maps, all images were created at the resolution of the $^{13}$CO map.
\section{Results}
\label{sec:result}
\subsection{Morphology}
\label{sec:morphology} 
\begin{figure*}
\begin{tabular}{lll}
\small \textbf{(a)}
\includegraphics[scale=0.408]{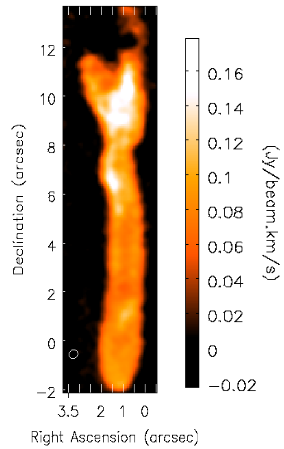}
\small \textbf{(b)}
\includegraphics[scale=0.403]{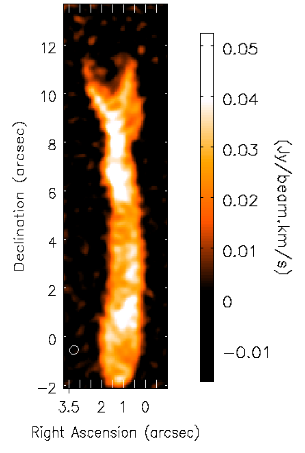}
&
\small \textbf{(c)}
\includegraphics[scale=0.405]{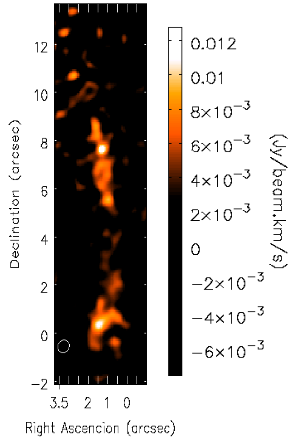}
\end{tabular}
\caption{CO integrated intensity maps of the C1 globule. The elliptical restoring beams (in arcsec $\times$ arcsec) are located at the bottom left of each map. \textbf{(a)}: $^{12}$CO with beam size $0.41\times0.35$, \textbf{(b)}:$^{13}$CO with beam size $0.42\times0.36$  and \textbf{(c)}: C$^{18}$O with beam size $0.43\times0.36$.}
\label{fig:integrated_CO}
\end{figure*}

\begin{figure*}
\includegraphics[scale=0.65]{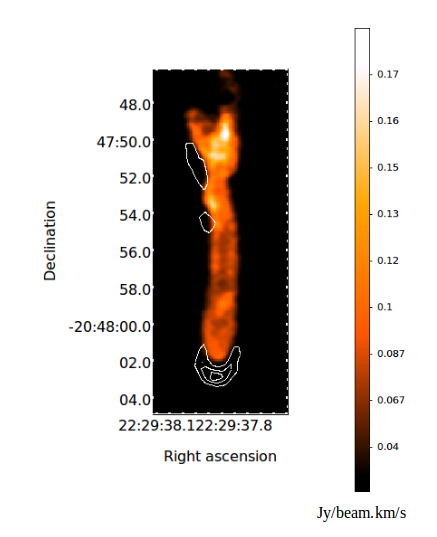}
\caption{$^{12}$CO integrated intensity map overlaid with \textit{HST} H$\alpha\,$+[N$\,$\textsc{ii}] contour map. The contours levels are at $0.76$, $1.53$, $2.31$ and $3.08$ in $10^{-22}\,\textmd{erg}\,\textmd{cm}^{-2}\,\textmd{s}^{-1}\,$\AA$^{-1}$. The molecular head is $\sim1^{\prime \prime}$ offset to the crescent-shaped head observed in the optical.}
\label{fig:overlay_HST}
\end{figure*}

CO emission in the cometary knots of the Helix nebula was first detected by \citet{huggins1992}. The molecular nature of the knots was then supported by the findings of \citet{huggins2002} who imaged the C1 globule in CO(1--0) with a velocity resolution of $0.2\,\textmd{km}\,\textmd{s}^{-1}$ and a spatial resolution of 3.6 arcsec. With a velocity resolution of $0.05\,\textmd{km}\,\textmd{s}^{-1}$ and an angular resolution of $0.39\,$arcsec, our data enables us to closely observe the molecular distribution of the C1 globule.
\\The integrated intensities of the $^{12}$CO and $^{13}$CO line emissions shown in Fig. \ref{fig:integrated_CO} confirm that the C1 knot is molecular, with strong emission from the head of the knot as well as from its extended tail. The structure of the globule is not well defined in the C$^{18}$O integrated intensity map as the emission is weak.
\\
\\The molecular head of the globule is ellipsoidal and is $\sim1^{\prime \prime}$ offset to the crescent-shaped ionized head observed in the optical (e.g. \citeauthor{o1996} \citeyear{o1996}) as shown in Fig. \ref{fig:overlay_HST}.
The tail is pointing away from the central star with a roughly cylindrical shape. We define the tail as the region of the knot starting at $\sim2^{\prime\prime}$ away from the tip of the head of the globule in CO.
\\We notice the presence of peak intensities at different positions in the knot in the $^{12}$CO and $^{13}$CO maps (Fig. \ref{fig:integrated_CO}). Some of the peaks are located close to the head, but peaks are also observed in the body, and strong emission is emitted in the tail region. At each peak, the direction of the tail seems to slightly deviate. This phenomenon is known as tail meandering. \citet{matsuura2009} reported that globules can present one or multiple peaks where a change of the direction of the tail can occur, which is in agreement with our observations. 
\\We find that the peak intensity of the emission seems to be higher in the tail than in the head of the knot. This is in contrast with the result of \citet{matsuura2007} who observed that the peak intensity decreases by a factor of 10 between the head and the tail in H$_2$. In addition, \citet{matsuura2009} noticed the existence of brightness gaps at the midpoint along the tail in different knots, which are also visible in our $^{12}$CO data, at $\sim+$0.5, $+\,$2.5 and $+\,5^{\prime\prime}$ (Fig. \ref{fig:integrated_CO}). In $^{13}$CO, the gap is observed in the head of the globule.
\subsection{Line intensity}
\begin{figure*} 
\centering
\includegraphics[scale=0.75]{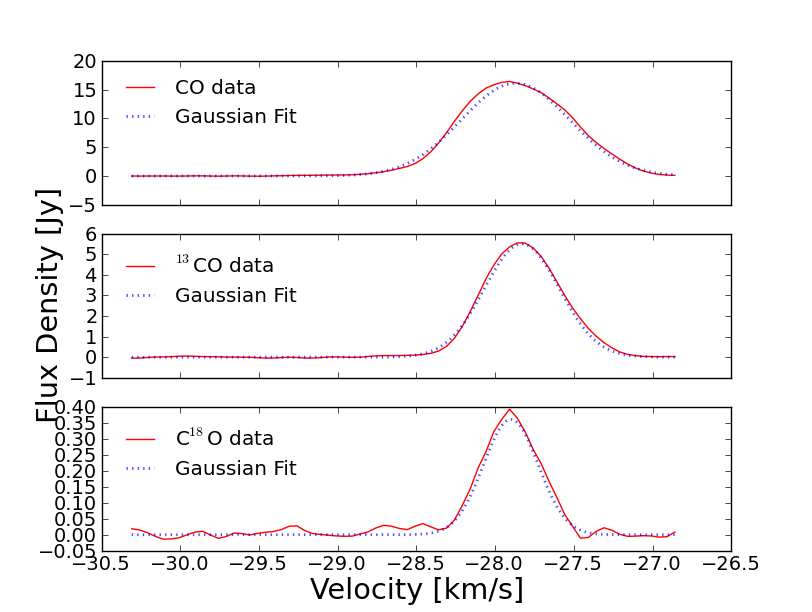}
\caption{Spectral profiles of the CO(2--1) lines with corresponding Gaussian fittings. The $^{12}$CO (top) Gaussian line is centred at $-27.87\,\textmd{km}\,\textmd{s}^{-1}$ with a peak intensity of $16.1\,\textmd{Jy}$ and a FWHM of $0.81\,\textmd{km}\,\textmd{s}^{-1}$. $^{13}$CO (middle), centred at $-27.83\,\textmd{km}\,\textmd{s}^{-1}$ with a peak intensity of $5.5\,\textmd{Jy}$, has a FWHM of $0.56\,\textmd{km}\,\textmd{s}^{-1}$. (bottom) shows the C$^{18}$O spectral profile  centred at $-27.91\,\textmd{km}\,\textmd{s}^{-1}$ with a peak value of $0.36\,\textmd{Jy}$ and a FWHM of $0.41\,\textmd{km}\,\textmd{s}^{-1}$. }
\label{fig:Spectral profile}
\end{figure*}

The integrated $^{12}$CO, $^{13}$CO and C$^{18}$O spectra and their respective Gaussian fittings are shown in Fig. \ref{fig:Spectral profile}. The $^{12}$CO line emission has a peak intensity of  16.1$\,\textmd{Jy}$, and a full width at half maximum (FWHM) of 0.81$\,\textmd{km}\,\textmd{s}^{-1}$. The $^{13}$CO line peaks at 5.5$\,\textmd{Jy}$, with a FWHM of 0.56$\,\textmd{km}\,\textmd{s}^{-1}$. The C$^{18}$O spectral profile presents a peak value of 0.36$\,\textmd{Jy}$ and a FWHM of 0.41$\,\textmd{km}\,\textmd{s}^{-1}$. The $^{12}$CO emission has a total flux of 14.4$\,\textmd{Jy}\,\textmd{km}\,\textmd{s}^{-1}$. The total $^{13}$CO and C$^{18}$O  fluxes are 3.5 and 0.18$\,\textmd{Jy}\,\textmd{km}\,\textmd{s}^{-1}$, respectively.  
\subsection{Optical depth}
\label{sec:optical depth}
\begin{figure*}
\includegraphics[width=0.92\textwidth, angle=0]{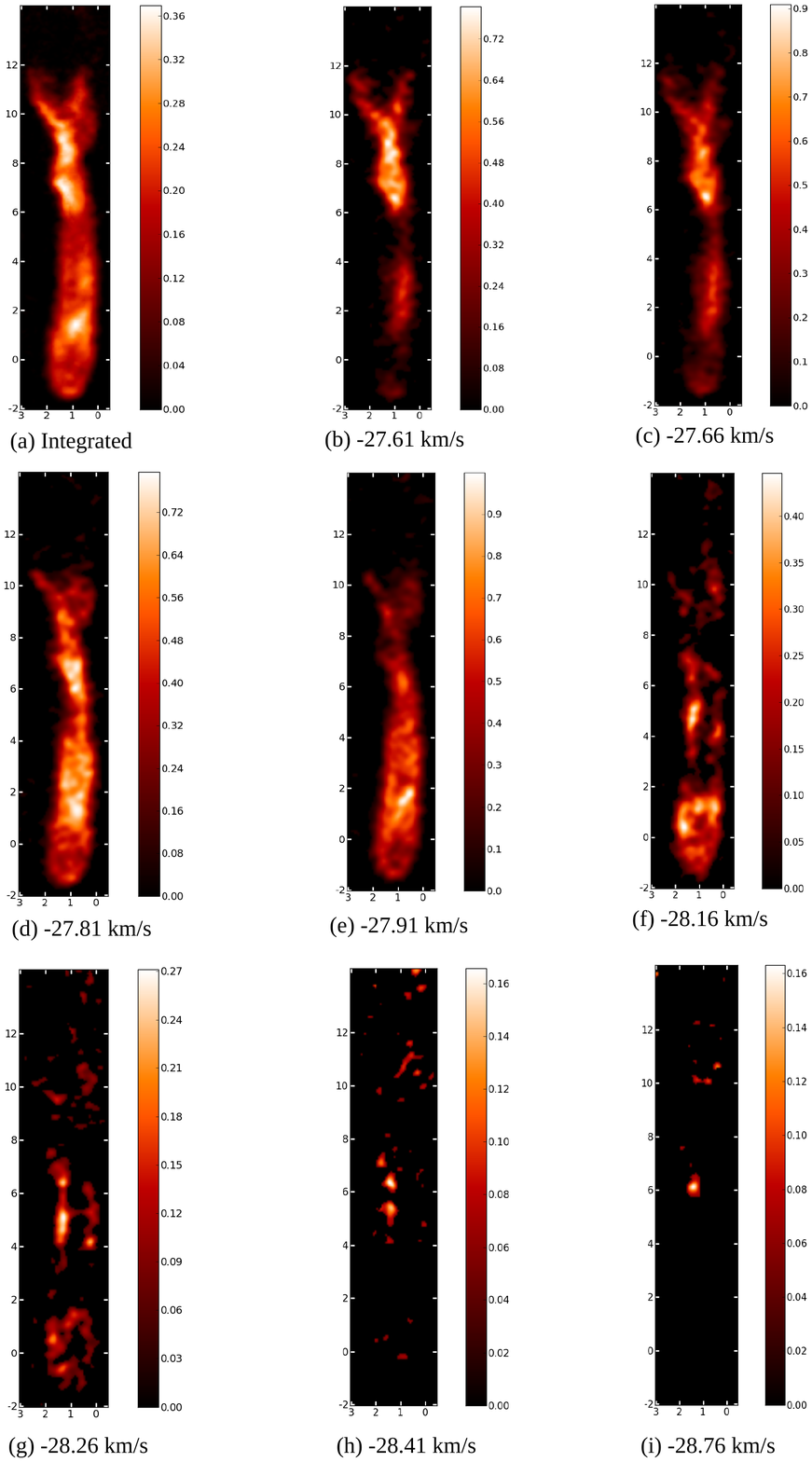}
\caption{ \textbf{(a)} is $^{13}$CO optical depth integrated map and \textbf{(b--i)} are $^{13}$CO optical depth channel maps.}
\label{fig:optical_depth}
\end{figure*}

Under the assumption of Local Thermodynamic Equilibrium (LTE), and assuming that $^{12}$CO, $^{13}$CO and C$^{18}$O have the same excitation temperature T$_\textmd{ex}$ and that $^{12}$CO is optically thick, we derive T$_\textmd{ex}$ from the  $^{12}$CO line using  \begin{equation}
    T_\textmd{ex}= 11.06 \, \Bigg\{ \textmd{ln}\, \bigg[ 1 + \frac{11.06}{T_\textmd{peak}^{12,2-1} + 0.19}\bigg]\Bigg\}^{-1} \, \textmd{K}
	\label{eq:temp}
\end{equation}
\\(e.g. \citeauthor{nishimura2015} \citeyear{nishimura2015}) where ${T_\textmd{peak}^{12,2-1}}=24.96\,\textmd{K}$ is the peak intensity of the $^{12}$CO(2--1) emission. We find $T_{\textmd{ex}}=\,30\,\textmd{K}$.
Masks were applied to the $^{13}$CO and C$^{18}$O integrated intensity maps in order to pick only the emission from the real source and avoid background noise, with cut-offs at $3\sigma_{13}$ and $2\sigma_{18}$ for $^{13}$CO and C$^{18}$O, respectively. The optical depths of $^{13}$CO and C$^{18}$O, $\tau_{13}$ and $\tau_{18}$, are given by \citep[e.g.][]{nishimura2015}
\begin{equation}
    \tau_{13}= - \textmd{ln} \Bigg\{1- \frac{T_\textmd{mb}^{13,2-1}(v)}{10.58} \Bigg[\frac{1}{\textmd{exp(10.58/$T_\textmd{ex}$)-1}}-0.02 \Bigg]^{-1}\Bigg\}
	\label{eq:opticaldepth}
\end{equation}
\begin{equation*}
    \tau_{18}= - \textmd{ln} \Bigg\{1- \frac{T_\textmd{mb}^{18,2-1}(v)}{10.54} \Bigg[\frac{1}{\textmd{exp(10.54/$T_\textmd{ex}$)-1}}-0.02 \Bigg]^{-1}\Bigg\}\,\textmd{,}
\end{equation*}
where $T_\textmd{mb}$ is the line temperature at a specific velocity $v$. Using the Rayleigh-Jeans law, the maps in units of intensity are converted to maps in units of brightness temperature with \citep[e.g.][]{thompson1986}
\begin{equation}
    T_\textmd{mb}= 1.222 \times10^{3} \frac{I}{\nu^2\,\theta_\textmd{maj}\,\theta_\textmd{min}}\, \textmd{K}\,\textmd{,}
    \label{eq:consersion}
\end{equation}
where $I$ is the intensity in mJy$\,$beam$^{-1}$, $\nu$ is the frequency in GHz. $\theta_\textmd{maj}$ and $\theta_\textmd{min}$ are the major and minor axes of the synthesised Gaussian beam. For the $^{13}$CO map, $\theta_\textmd{maj}=0.42\,$arcsec and $\theta_\textmd{min}=0.36\,$arcsec, whilst for the C$^{18}$O map $\theta_\textmd{maj}=0.43\,$arcsec and $\theta_\textmd{min}=0.36\,$arcsec.\\
We find a maximum optical depth of $0.99$ and $0.16$ for $^{13}$CO and C$^{18}$O, respectively. 
Using the observed $^{12}$CO/$^{13}$CO ratio of $\sim$10 found by \cite{bachiller1997} for the entire nebula, we estimate the $^{12}$CO optical depth to be up to $\sim$10.
\citet{huggins1992} assumed that $^{12}$CO in the C1 globule is optically thin. With the higher resolution achieved by ALMA, we find that $^{13}$CO is mostly optically thin and C$^{18}$O is always optically thin, while the $^{12}$CO emission is highly optically thick.\\
Figure \ref{fig:optical_depth} shows the $\tau_{13}$ map integrated over all the channels, as well as the $^{13}$CO  optical depth channel maps of the C1 globule. Assuming the previously mentioned $^{12}$CO and $^{13}$CO ratio of $\sim\,$10, we look for optically thin $^{12}$CO regions where $\tau_{13}<0.1$. 
\\We observe that the optical depth is not the same along the knot. The integrated optical depth map (Fig. \ref{fig:optical_depth}\textit{a}) shows that the inner region of the globule is optically thick in $^{12}$CO, and the highest optical depths are located at declinations from $+0\textmd{ to}+3^{\prime\prime}$ and $+\,6\textmd{ to}+10^{\prime\prime}$. The extended tail (at declination $>+10^{\prime\prime}$) shows the lowest optical depth in $^{13}$CO ($\sim$0.1). The channel maps show how the $^{13}$CO optical depth varies along the globule at different velocities. In Fig. \ref{fig:optical_depth}\textit{b} and \ref{fig:optical_depth}\textit{c}, the tail presents high optical depth, whereas the head appears less optically thick. The region at declinations between $+1^{\prime\prime}$ and $+4^{\prime\prime}$ is always optically thick. In the tail region, the $^{13}$CO optical depth reaches up to $0.9$ at a velocity of $-27.65\,\textmd{km}\,\textmd{s}^{-1}$ and a declination of $\sim +6^{\prime\prime}$ (Fig. \ref{fig:optical_depth}\textit{c}). In Fig. \ref{fig:optical_depth}\textit{d}, the entire inner part of the body is optically thick. The maximum optical depth $\tau_{13\textmd{max}}=0.99$ is observed close to the head at a declination of $+1.7^{\prime\prime}$ and at a velocity of $-27.90\,\textmd{km}\,\textmd{s}^{-1}$ (Fig. \ref{fig:optical_depth}\textit{e}). The head of the globule is optically thick in Fig. \ref{fig:optical_depth}\textit{f} while the higher part of the tail seems to be optically thin. In Fig. \ref{fig:optical_depth}\textit{g}, a relatively high optical depth value is observed at a declination of $+6^{\prime\prime}$, which could potentially represent a separate knot. This feature is observed from Fig. \ref{fig:optical_depth}\textit{f} to Fig. \ref{fig:optical_depth}\textit{i}. 
\section{Isotopic ratios}
\label{sec:ratio}
\begin{figure*}
\begin{tabular}{ll}
\small \textbf{(a)}
\includegraphics[scale=0.6153]{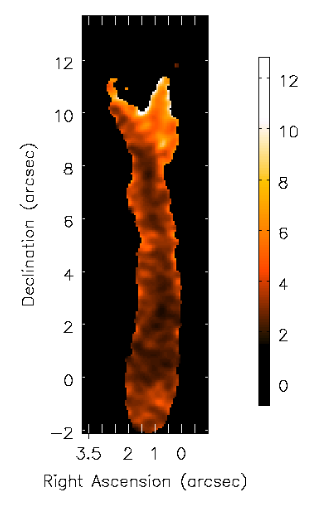}
&
\small \textbf{(b)}
\includegraphics[scale=0.5]{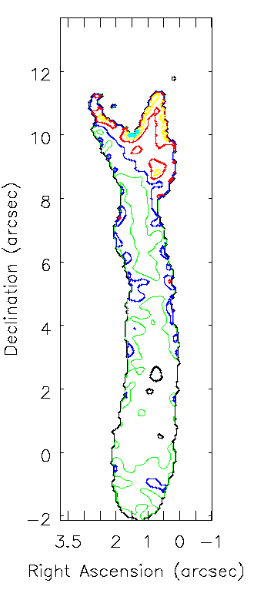}
\end{tabular}
\caption{ $^{12}$CO/$^{13}$CO ratio map integrated over the velocity axis, in units of flux. The ratio increases towards the tail of the globule. The central region of the knot presents very low ratio ($< 4$). The extended part of the tail has a ratio of $\sim$6, reaching up to $>$10 in the edges. The contours in \textbf{(b)} have ratio values of 2 (black), 3 (green), 4 (blue), 6 (red), 8 (yellow) and 11 (cyan).}
\label{fig:ratio_12_13}
\end{figure*}

\begin{figure*}
\includegraphics[scale=0.6]{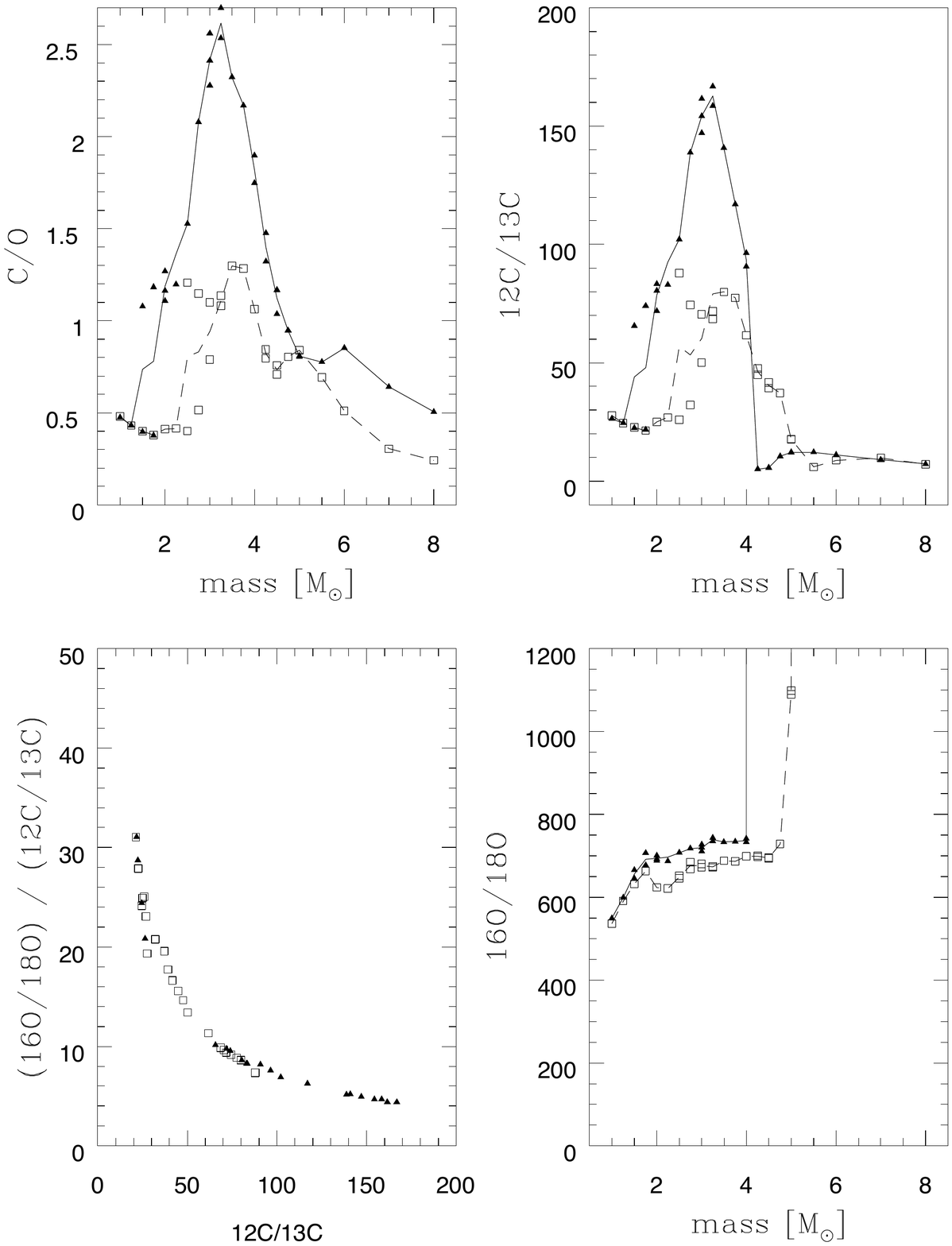}
\caption{Predicted isotopic and elemental abundance ratios at the end of the AGB, as function of initial stellar mass, taken from \citet{Karakas2016}.}
\label{fig:model ratio}
\end{figure*}

To find the $^{12}$CO/$^{13}$CO ratio of the globule, we first regrid the $^{13}$CO integrated intensity map so that the corresponding pixel coordinates match to those of the $^{12}$CO map, with both maps in units of flux. To minimise the effects of resolution on the division in the border region, we assign the same restoring beam size for both maps. We create a binary mask that picks only the regions where the $^{13}$CO emission is above the $3\sigma_{13}$ level. The created mask is then applied to both the $^{12}$CO and $^{13}$CO integrated intensity maps in order to avoid thermal noise in the ratio calculation. We find that the $^{12}$CO/$^{13}$CO line ratio is not uniform throughout the cometary knot. The edges and the core of the globule present different values of $^{12}$CO/$^{13}$CO, as illustrated in Fig. \ref{fig:ratio_12_13}. In the inner parts of the head and the tail, the observed isotopic line ratio is $<$4, with particularly low values ($\sim$2) in the central part. The ratio tends to be higher in the edges of the tail region, reaching up to $\sim$11. Given the age and the physical conditions inside the Helix nebula, the underlying isotopic ratio $^{12}$C/$^{13}$C is expected to be uniform. 
\\
\\An isotope-selective-photo-dissociation mechanism could be a reason for the variable $^{12}$CO/$^{13}$CO line ratio \citep{saberi2019}. According to \citet{szHucs2014}, the photo-dissociation process in a molecular cloud is closely related to the CO density of the region in the cloud. At relatively low-density, $^{12}$CO is self-shielded  from the UV radiation from the CS because the rates at which the photo-dissociation and the $^{12}$CO production processes occur are similar. On the other hand, at those density levels, $^{13}$CO is not as effectively shielded due to its lower abundance and its absorption properties. In higher density regions, both $^{12}$CO and $^{13}$CO are effectively shielded yielding to a decrease in the ratio. The selective photo-dissociation is dominant mainly in low density regions such as in the extended tail of the globule, and could be the reason behind the observed high values of the $^{12}$CO/$^{13}$CO line ratio ($\sim$11). However, it does not explain the low values of the ratios. These are more likely due to optical depth effects.\\
\\In section \ref{sec:optical depth}, we find that $^{13}$CO is mostly optically thin, whereas $^{12}$CO is highly optically thick. The non-uniformity of the optical depth of the  $^{13}$CO lines along the globule as seen in Fig. \ref{fig:optical_depth} is likely to be the main reason for the non-constant observed isotopic ratio. The lowest $^{12}$CO/$^{13}$CO line ratio observed in the centre of the globule corresponds to the highest optical depth values, i.e. optically thick $^{12}$CO region. However, the extended tail of the globule is likely to be optically thin in $^{12}$CO. The emission from this optically thin region would therefore give a better estimate of the true $^{12}$CO/$^{13}$CO ratio. Using Fig. \ref{fig:ratio_12_13}, we find that the $^{12}$CO/$^{13}$CO intensity ratio corresponding to this region optically thin in $^{12}$CO is around $11\pm1$. We take this value of $\sim$11 as the $^{12}$CO/$^{13}$CO intensity ratio of the whole globule corrected for optical depth. 
For optically thin lines, a first estimate of the $^{12}$C/$^{13}$C isotopic ratio can be  calculated using the correlation between the $^{12}$C/$^{13}$C and the $^{12}$CO/$^{13}$CO intensity ratio given by
\begin{align}
    ^{12}\textmd{C}/^{13}\textmd{C}=\frac{I(^{12}\textmd{CO($J\,\to\,J$--1)})}{I(^{13}\textmd{CO($J\,\to\, J$--1)})} \times \bigg (\frac{\nu_{^{12}\textmd{CO($J\,\to\, J$--1)}}}{\nu_{^{13}\textmd{CO($J\,\to\, J$--1)}}} \bigg )^{-3}\textmd{ , }
    \label{eq:isotopic23}
\end{align}
where $I(^{12}\textmd{CO})$ and $I(^{13}\textmd{CO})$ are the intensities of the $^{12}$CO and $^{13}$CO lines, respectively. We use the derived optical depth corrected $^{12}$CO/$^{13}$CO intensity ratio of  $\sim$11 as the line intensity ratio and obtain $^{12}$C/$^{13}$C $\sim$10. 
\\
\\We follow the same procedure to determine the C$^{16}$O/C$^{18}$O ratio. The observed ratio varies from $5$, in the inner parts of the globule, to $>20$ in the outer regions. This C$^{16}$O/C$^{18}$O intensity ratio can be suppressed by optical depth effects. As $^{13}$CO and C$^{18}$O are both mostly optically thin, we use the $^{13}$CO/C$^{18}$O ratio to derive the optical depth-corrected C$^{16}$O/C$^{18}$O ratio. We find a mean $^{13}$CO/C$^{18}$O value of $\sim$12. As the $^{12}$CO/$^{13}$CO line ratio is $\sim11$, the C$^{16}$O/C$^{18}$O line ratio is therefore estimated to be $\sim$132. To obtain the $^{16}$O/$^{18}$O isotopic ratio, we use \citep[e.g.][]{nutte2017}
\begin{align}
     ^{16}\textmd{O}/^{18}\textmd{O}=\frac{I(\textmd{C}^{16}\textmd{O($J\,\to\, J$--1)})}{I(\textmd{C}^{18}\textmd{O($J\,\to\, J$--1)})} \times \bigg (\frac{\nu_{\textmd{C}^{16}\textmd{O($J\,\to\, J$--1)}}}{\nu_{\textmd{C}^{18}\textmd{O($J\,\to\, J$--1)}}} \bigg )^{-3} \textmd{ , }
    \label{eq:isotopic68}
\end{align}
where $I($C$^{16}\textmd{O})$ and $I($C$^{18}\textmd{O})$ are the intensities of the C$^{16}$O and C$^{18}$O lines, respectively. We find $^{16}$O/$^{18}$O$\simeq$115.
\\The $^{13}$CO/C$^{18}$O line ratio may be underestimated: with a maximum optical depth of $\tau_{13}=0.99$, a part of the real $^{13}$CO emission can be undetected. However, we assumed a constant excitation temperature which may have an effect on the optical depth calculation. The emission detected is from the coldest optically thick $^{12}$CO line. The optically thin  $^{13}$CO and C$^{18}$O emission could trace warmer gas, with a higher limit of 185$\,$K for the C$^{18}$O line. Increasing the value of $T_{\textmd{ex}}$ in Eq. \ref{eq:temp} would give lower values of the $^{13}$CO and C$^{18}$O optical depths. With an excitation temperature of 60$\,$K, we obtain a maximum $^{13}$CO optical depth of 0.34.  
Furthermore, it is important to point out that the ratios obtained are from potentially $^{12}$CO optically thin regions, based on the assumption that the optical depth of $^{12}$CO is higher than the optical depth of $^{13}$CO by a factor of 10 (section \ref{sec:optical depth}).
\\
\\
\\The derived isotopic ratios depend on the optical depth of the $^{12}$CO line which is not well constrained.  The AGB models of \citet{Karakas2016} are used to compare these ratios with what is predicted by those models. This is depicted in Fig. \ref{fig:model ratio} which shows the surface abundance ratios at the end of the AGB evolution. The values prior to the last thermal pulse only differ slightly from these values. Models are shown for solar metallicity (Z=0.014: triangles and drawn lines) and for supersolar metallicity (Z=0.03: square and dashed lines). \citet{Karakas2016} present models with different mixing assumptions. This is the reason that at the same stellar mass, several values may be shown. The drawn and dashed line present an ensemble average, but the points show the spread at each mass.
\\An important constraint is the $^{16}$O/$^{18}$O ratio. In the models, for initial stellar masses above 4-5$\,$M$_\odot$, this ratio becomes very large ($>10^6$) because $^{18}$O is burned. The C$^{18}$O detection shows that the progenitor mass must be below this range.
\\The second constraint comes from the suggestion that the dust in the helix is carbonaceous \citep{van2015}. The chemical nature of the Helix nebula has been a subject of debate in the literature. \citet{henry1999} found that oxygen is more abundant than carbon in the Helix nebula using atomic emission lines. The detection of carbon-rich molecules in the nebula, however, led to the conclusion of its carbon molecular nature (\citeauthor{cox1998} \citeyear{cox1998}, \citeauthor{tenenbaum2009} \citeyear{tenenbaum2009}). In recent studies (e.g. \citeauthor{van2015} \citeyear{van2015}), the Helix nebula is treated as a carbon-rich nebula based on the emissivity index of the dust emission.  Carbonaceous dust requires that C/O\,$> 1$. As shown in Fig. \ref{fig:model ratio}, for the \citet{Karakas2016} models this happens for initial stellar masses between 1.5 and 4.5$\,$M$_\odot$ at solar metallicity.
\\Within this mass range, the isotopic ratios found here are not reproduced by the models. The $^{12}$C/$^{13}$C ratio is above 50 in all models, due to the fact that the $^{12}$C dredge-up is needed to form a carbon star. Outside the carbon star range, the $^{12}$C/$^{13}$C$\sim 10$ is found for high masses but this mass range is excluded by the $^{18}$O detection. Elsewhere, the $^{16}$O/$^{18}$O ratio is always around 700, rather than 115 as found here.
\\An important diagnostic is the ratio between $^{16}$O/$^{18}$O and $^{12}$C/$^{13}$C, which is around 10 in our data. Both parts of this ratio depend in the same way on the $^{12}$CO line, and as long as both the $^{13}$CO and C$^{18}$O line are optically thin, this ratio should come out correct. If the $^{13}$CO line is optically thick, the ratio becomes a lower limit.  Figure \ref{fig:model ratio} shows the models for this ratio against $^{12}$C/$^{13}$C. For the observed value, $^{12}$C/$^{13}$C$\sim 50$, while for higher values it reduces. Comparing this to the models gives a possible mass around 2$\,$M$_\odot$. There is a second solution around 4$\,$M$_\odot$. Because of the steep initial mass function, there is some preference for the lower mass.
\\For this result to be correct, the $^{12}$C/$^{13}$C ratio derived here would need to be underestimated, and therefore the $^{12}$CO line would need to have a higher optical depth. We cannot confirm this with the available data.
We can also compare these predictions with observations of the $^{12}$C/$^{13}$C isotopic ratios of presolar grains in \citet{zinner2014}. The optical depth-corrected $^{12}$C/$^{13}$C isotopic ratio of the C1 globule is lower than the $^{12}$C/$^{13}$C ratios of mainstream (20<$^{12}$C/$^{13}$C$<$100) and type Z silicon carbide grains ($^{12}$C/$^{13}$C$<$100), which originate from AGB carbon stars \citep{zinner2014}. The presolar grains show isotopic ratios that agree with stellar models.
\\We can also compare with observations of AGB stars. The derived $^{12}$C/$^{13}$C ratio is within the range of the $^{12}$C/$^{13}$C ratios for different SC-type AGB carbon stars (C/O$\sim 1$) in \citet{abia2017}, which range from 5 to 35. However, the obtained $^{16}$O/$^{18}$O ratio of 115 in the Helix is much lower than the $^{16}$O/$^{18}$O ratios in \citet{abia2017} for different SC-type AGB carbon stars, ranging from $400$ to $1250$. 
\\This discussion shows that the $^{13}$C$^{16}$O/$^{12}$C$^{18}$O is an important observational constraint in the comparison with the models. In the case of the Helix, the comparison raises the possibility that the optical depth of the $^{12}$CO line has been underestimated.
\section{Mass}
\subsection{Molecular mass}
\label{sec:mass}
\begin{figure*}
\includegraphics[scale=0.55]{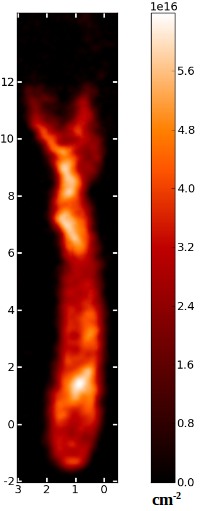}
\caption{ $^{13}$CO Column density map of the C1 globule. The axes are the same as in Fig. \ref{fig:optical_depth}. The tail is denser than the head.}
\label{fig:columndensity}
\end{figure*}

We use the CO lines as H$_2$ tracers to estimate the molecular mass of the C1 globule.
\\The column densities  of $^{13}$CO and C$^{18}$O, N($^{13}$CO) and N(C$^{18}$O), are calculated using the column density in the upper state $N_\textmd{$J=2$}$ \citep[e.g.][]{nishimura2015}
\begin{equation}
    N_\textmd{$J=2$}^\textmd{$^{13}$CO}= 1.65 \times 10^{16} \Bigg[ \textmd{exp} \Bigg(\frac{10.58}{T_\textmd{ex}}\Bigg)-1\Bigg]^{-1}\int\tau_{13}(v)\,dv \, \textmd{cm$^{-2}$}
	\label{eq:columndensity}
\end{equation}
\begin{equation*}
 N_\textmd{$J=2$}^\textmd{C$^{18}$O}= 1.64 \times 10^{16} \Bigg[ \textmd{exp} \Bigg(\frac{10.54}{T_\textmd{ex}}\Bigg)-1\Bigg]^{-1}\int\tau_{18}(v)\,dv \, \textmd{cm$^{-2}$}
\end{equation*}
where $T_\textmd{ex}=30\,\textmd{K}$ is the excitation temperature, $v$ is the velocity centre and $dv=0.05 $ $\textmd{km}\,\textmd{s}^{-1}$ is the velocity width. $\tau_{13}$ and $\tau_{18}$ are the optical depths of $^{13}$CO and C$^{18}$O calculated in section \ref{sec:optical depth}. The total column density is obtained using \citep{nishimura2015}
\begin{equation}
    N(\textmd{$^{13}$CO})= N_\textmd{$J=2$}^\textmd{$^{13}$CO}\, \frac{Z}{2\,J+1}\,\textmd{exp}\bigg[\frac{h\,B_0\,J(J+1)}{k\,T_\textmd{ex}} \bigg]\, \textmd{cm$^{-2}$}\, \textmd{,}
	\label{eq:columndensity total 13co}
\end{equation}
\begin{equation*}
    N(\textmd{C$^{18}$O})= N_\textmd{$J=2$}^\textmd{C$^{18}$O}\, \frac{Z}{2\,J+1}\,\textmd{exp}\bigg[\frac{h\,B_0\,J(J+1)}{k\,T_\textmd{ex}} \bigg]\, \textmd{cm$^{-2}$}\, \textmd{.}
	\label{eq:columndensity total c18o}
\end{equation*}
$N_\textmd{$J=2$}^\textmd{$^{13}$CO}$ and $N_\textmd{$J=2$}^\textmd{C$^{18}$O}$ are the column density in the state $J=2$ for $^{13}$CO and C$^{18}$O derived in equation (\ref{eq:columndensity}), respectively. $B_0$ is the rotational constant, where $B_0=5.51 \times 10^{10} \,$s$^{-1}$ for $^{13}$CO and $B_0=5.49 \times 10^{10} \,$s$^{-1}$ for C$^{18}$O \citep{nishimura2015}. $h$ and $k$ are the Planck and Boltzmann constant, respectively. $T_\textmd{ex}$ is the excitation temperature obtained in equation (\ref{eq:temp}). $Z$ is the partition function given by \citep{nishimura2015}
\begin{equation}
    Z=\sum_{J=0}^{\infty} (2\,J+1)\,\,\textmd{exp} \bigg[ -\frac{h\,B_0\,J(J+1)}{k\,T_\textmd{ex}} \bigg ] 
	\label{eq:partition function}
\end{equation}
which can be approximated as follows for diatomic linear molecules \citep{Mangum2015}
\begin{equation}
    Z \simeq \frac{k\,T}{h\,B_0} + \frac{1}{3}+\frac{1}{15} \bigg( \frac{h\,B_0}{k\,T} \bigg) + \frac{4}{315} \bigg ( \frac{h\,B_0}{k\,T} \bigg)^2 + \frac{1}{315} \bigg ( \frac{h\,B_0}{k\,T} \bigg)^3 + ... \, \textmd{.}
	\label{eq:partition function approx}
\end{equation}
Equation (\ref{eq:partition function approx}) is accurate to $<$1 percent for $T_\textmd{ex}>\,$2$\,$K, even with the first two terms only \citep{Mangum2015}.\\
\\
The derived  column densities $N(\textmd{$^{13}$CO})$ and $N(\textmd{C$^{18}$O})$ are converted to $^{12}$CO column densities by multiplying them with the corresponding isotopic ratios corrected for optical depth derived in section \ref{sec:ratio}. We obtain the two-dimensional $^{12}$CO column density distributions, N($^{12}$CO), integrated over all velocity channels. 
\\We use the distance to the Helix nebula of 201$\pm 4\,$pc derived from the parallax of the Helix measured by \citet{gaia} \citep[\textit{Gaia} mission:][]{gaia2018b,luri2018}, and assume a CO/H$_2$ ratio of 6$\,\times 10^{-4}$ based on the CO/H abundance ratio of $\sim$3$\,\times 10^{-4}$ in the Helix nebula in \citet{huggins1992,huggins2002}. This assumed CO/H$_2$ ratio is in agreement with the ratio found by \citet{debeck2010} for S-type stars.  We obtain a molecular mass of $1.8\,\times10^{-4}\,$\(\textup{M}_\odot\) from $^{13}$CO and of $1.5\times 10^{-4}\,$\(\textup{M}_\odot\) from the C$^{18}$O emission. \\
\\We look into the distribution of the mass in the knot using the head and tail definition given in section \ref{sec:morphology}. We find that the tail has a mass of $1.7\,\times 10^{-4}\,$\(\textup{M}_\odot\) from $^{13}$CO and $1.5\,\times10^{-4}\,$\(\textup{M}_\odot\) from C$^{18}$O, whereas the head only accounts for $1.3\,\times10^{-5}$ and $7.7\,\times10^{-6}\,$\(\textup{M}_\odot\) from $^{13}$CO and C$^{18}$O, respectively. This means that the majority of the CO in the knot is carried by the tail. The  $^{13}$CO column density distribution of the head and the tail of the C1 globule is illustrated in Fig. \ref{fig:columndensity} and shows that the tail is denser than the head. \\
\\These derived masses are $>$9 times higher than the mass found by \citeauthor{huggins2002} (\citeyear{huggins2002}, \citeyear{huggins1992}) for C1 using CO(1--0) observations, and \citet{young1997} for different knots using CO(2--1), ranging from 5$\,\times10^{-6}$ to 2$\,\times10^{-5}\,$\(\textup{M}_\odot\). The difference between our result and the masses obtained in those previous studies is mainly due to optical depth. \citet{huggins1992,huggins2002} calculated the molecular mass of the C1 globule from CO(1--0) emission assuming that $^{12}$CO is optically thin. However, with an optically thick $^{12}$CO and making use of its optically thin isotopes, $^{13}$CO and C$^{18}$O, we derive the molecular mass considering the $^{12}$CO column density corrected for optical depth. Without optical depth correction, the molecular mass would be $\sim$6$\,\times10^{-5}\,$\(\textup{M}_\odot\), which is closer to the mass found by \citet{huggins2002}. In addition, the mass derived from some of these prior studies are obtained from a beam-averaged column density from unresolved knots which led to an underestimation due to beam dilution.   
\\Uncertainties lie in the CO/H$_2$ conversion factor and the LTE assumption, which have been found to estimate the molecular mass with a discrepancy of up to $\sim30$ percent \citep{szHucs2016}. \\
\\With $23\,000$ globules in the Helix nebula \citep{meixner2005} with individual masses of $\sim$2$\,\times10^{-4}\,$\(\textup{M}_\odot\), the total mass of the globules would give a value much higher than the mass of the entire nebula, M$_{\textmd{Helix}}=\,\sim$1.5$\,$\(\textup{M}_\odot\) \citep{speck2002}. This could mean that the Helix nebula is, in fact, a combination of condensed globules. However, an overestimation of the number of the knots is possible, though \citet{matsuura2009} found a  higher number of globules by almost a factor of 2. Furthermore, globules do not necessarily have the same mass. As seen in Fig. 3 in \citet{meaburn1998} where C1 is designated 1, this particular globule is relatively large, hence more massive than some of its companions. In addition, our results show that the mass of C1 is mainly found in the tail. As only the knots in the inner part like C1 are observed to exhibit distinct tails (e.g. \citeauthor{matsuura2009} \citeyear{matsuura2009}), the knots in the outer regions could have a much lower mass than those located closer to the CS.\\
\\The derived physical parameters of the C1 globule are presented in Table \ref{tab:table1}. 
\begin{table*}
	\centering
	\caption{Derived properties of the C1 globule.}
	\label{tab:table1}
	\begin{tabular}{lccccccccccr} 
		\hline
         & Size &  Flux & Flux density & Mass of the head & Mass of the tail & Total mass& \\
		 & [arcsec $\times$ arcsec] & [Jy$\,\textmd{km}\,\textmd{s}^{-1}$] & [mJy] & [\(\textup{M}_\odot\)] & [\(\textup{M}_\odot\)] & [\(\textup{M}_\odot\)]\\
		\hline
		CO  &  $\sim2\,\times\,16$ & 14.4 & -&-&-&-\\
		$^{13}$CO  & $\sim2\,\times\,14$ & $3.5$&- &$1.3\times10^{-5}$ & $1.7\times10^{-4}$ & $1.8\times10^{-4}$\\
		C$^{18}$O & $\sim 1-2\,\times\,11$& $0.18$&- &$7.7\times10^{-6}$ & $1.5\times10^{-4}$ & $1.5\times10^{-4}$\\
        Dust & -& -& $0.33$ &$2.7\times10^{-6}$&$2.8\times10^{-6}$&  $5.5\times10^{-6}$ \\
		\hline
	\end{tabular}
\end{table*}

\subsection{Gas mass from dust observation}
\label{sec:dust}
\begin{figure*}
\includegraphics[scale=0.6]{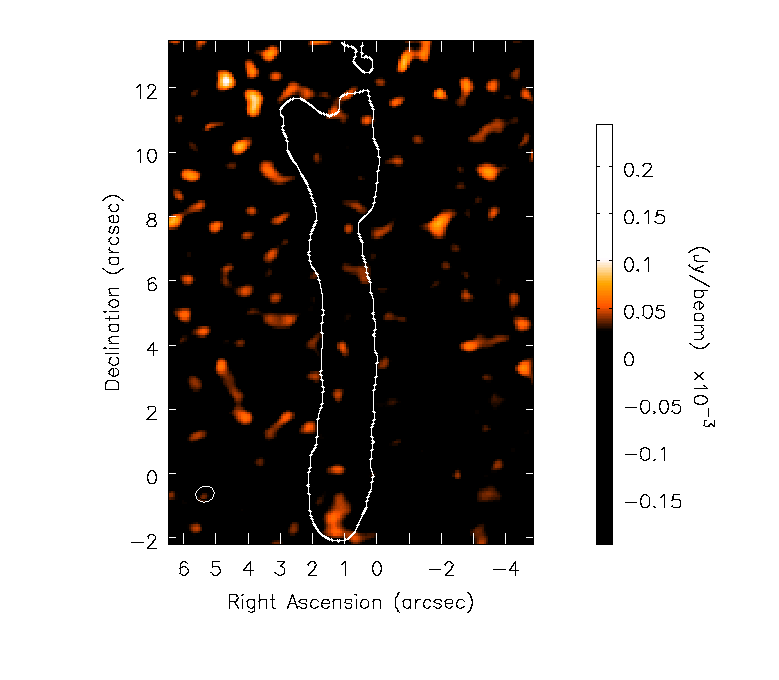}
\caption{Dust continuum map with $^{12}$CO contour at a level of $20\,\textmd{mJy}\,\textmd{beam}^{-1}$. The restoring beam (in  $\textmd{arcsec}\times\textmd{arcsec}$) is at the bottom left of the map with a size of $0.40 \times 0.34$.}
\label{fig:Dust_map}
\end{figure*}

Figure \ref{fig:Dust_map} shows the $1.2\,\textup{mm}$ continuum map of the C1 globule. Free-free emission  is not expected to be important in the knot, so the continuum emission is entirely associated with dust, with a total flux density of $(3.3\pm 0.2)\times10^{-4}${Jy}. There is no zero-spacing problem for the narrow tail in the interferometric data: all flux is in principle detected. We do not consider emission below  2$\sigma$ ($\sigma=20\,\mu$Jy$\,$beam$^{-1}$). Figure \ref{fig:Dust_map} shows that dust in the tail is weaker compared to the observed CO emission. Dust is only detected at a few locations throughout the tail of the knot whereas the head region seems to present a relatively concentrated dust emission, accounting for $\sim50$ percent of the total flux density.\\
\\We calculate the gas mass of the globule using the measured dust emission. On the assumption that grains in the nebula are mainly amorphous carbon, \citet{van2015} found that the temperature of the dust in the Helix nebula varies from $22$ to $42\,\textmd{K}$ with a mean value of $30.8\pm1.4\,\textmd{K}$. We use
\begin{equation}
    M_{\textmd{gas}}=\frac{10^{-26}}{T_{\textmd{dust}} \,\chi_\textmd{d}} \frac{S_\nu\, D^2\,\lambda^2}{2\,\gamma K_\textmd{d}\, k_\textmd{B}} \frac{1}{2} 10^{-33} \, \textup{M}_\odot\
	\label{eq:Mgas}
\end{equation}
to find the mass of the gas (e.g. \citeauthor{evans2004} \citeyear{evans2004}), assuming that the emission is under the Rayleigh-Jeans approximation. The parameters used are as follows. $T_\textmd{dust}$ is the mean temperature of the dust, found by \cite{van2015}. The temperature of the dust in the Helix nebula is similar to the excitation temperature of the CO lines obtained in section \ref{sec:mass}. This is indicative that molecular gas and dust coexist inside the globule. $K_\textmd{d}$  corresponds to the absorption coefficient. We use $K_\textmd{d}=20\,\,\textmd{cm}^2\,\textmd{g}^{-1}$ taken from \citet{mennella1998} for amorphous carbon dust grain at $1\,\textmd{mm}$ and at $20\,\textmd{K}$, which is the closest to our data specifications. $\chi_\textmd{d}$ is the dust--to--gas mass ratio. Its value varies in the literature, from $\chi_\textmd{d}\simeq1\textmd{:}55$ in \citet{o1998} and \citet{henry1999} to $\chi_\textmd{d}=1\textmd{:}150$ in \citet{sodroski1994} and \citet{o2005}.  Here we use $\chi_\textmd{d}=1\textmd{:}100$ as in \citet{meaburn1992}. $D$ is the distance to the Helix nebula in cm, $S_v$ is the flux density in Jy and $\lambda$ represents the wavelength in m. $k_\textmd{B}$ is the Boltzmann constant and $\gamma$ is a constant that depends on the optical thickness of the dust shell. $\gamma=1$ for an optically thin dust shell, while $\gamma=2$ if the shell is optically thick. At this wavelength, dust emission is optically thin. \\
\\We obtain a dust mass of 5.5$\,\times10^{-8}\,$\(\textup{M}_\odot\) and a gas mass of 5.5$\,\times10^{-6}\,$\(\textup{M}_\odot\). This gas mass is $\sim$2 times lower than the globule mass in \citet{meaburn1992} and in \citet{meaburn1998} ($\sim10^{-5}$ \(\textup{M}_\odot\)) from dust observation which took into account the emission in the head of the globule only. The masses of the knot derived from the CO lines (section \ref{sec:mass}) and dust emission differ by a factor of >$\,$25. Figure \ref{fig:Dust_map} suggests that dust in the tail is much weaker than the observed CO, which is reflected in the mass derived from dust emission. The obtained gas mass is on the order of the molecular mass of the head of the globule that we derived from C$^{18}$O. As the detected dust emission is weak, the 2$\sigma$ cut-off could result in an underestimation of the total gas mass.\\
\\The value of the dust--to--gas mass ratio $\chi_\textmd{d}$ in the Helix nebula is uncertain. With a dust--to--gas mass ratio of $1\textmd{:}150$ in the Helix (\citeauthor{sodroski1994} \citeyear{sodroski1994}, \citeauthor{o2005} \citeyear{o2005}), we find a gas mass of 8.3$\,\times10^{-6}$ \(\textup{M}_\odot\). With the commonly used value of dust--to--gas mass ratio for the Helix nebula $\chi_\textmd{d}\simeq1\textmd{:}55$ (\citeauthor{o1998} \citeyear{o1998}, \citeauthor{henry1999} \citeyear{henry1999}), the corresponding gas mass is 3$\,\times10^{-6}$ \(\textup{M}_\odot\). 
\\Non-negligible uncertainties lie in  the absorption coefficient. According to \citet{mennella1998}, the absorption coefficient decreases as the wavelength increases. The value of $K_\textmd{d}$ that we used to calculate the dust mass is for a 1$\,$mm-wave at 20$\,$K, so, a more accurate value of $K_\textmd{d}$ corresponding to the observed 1.2$\,$mm dust emission at $30\,$K would give a higher gas mass. \\
\subsection{Clumping}
In section \ref{sec:ratio}, we mentioned the possibility of a non-constant excitation temperature. With the resolution of our ALMA data, we assumed a smooth CO distribution, but the existence of unresolved clumps with different temperatures and different optical depths throughout the globule could be possible. The presence of higher density molecular clumps that are smaller than the beam size in the knot would have led to an overestimation of the molecular mass, as the brightness measured is averaged over the beam. 
Knowing the size and the density of those clumps would give a better estimate of the molecular mass of the knot. With a beam dilution effect of $\sim$10 due to high density clumps, the molecular mass after beam dilution correction would be similar to the gas mass of C1 and other knots from dust emission derived by \citet{meaburn1992,meaburn1998} and to its molecular mass in previous studies \citep[e.g.][]{young1997,meaburn1998,huggins2002}. Clumps in the C1 globule would also affect the calculation of the optical depth and the isotopic ratios in the globule, and therefore, influence the determination of the chemical nature of the Helix nebula. However, \citet{meaburn1998} reported a uniform H$_2$ density throughout the head of the C1 globule. Higher angular resolution data, in the optically thin C$^{18}$O line in particular, are needed in order to investigate the likelihood of clumping in the C1 knot. 
\section{Origin and evolution}
\label{sec:origin}
The processes responsible for the origin and evolution of the cometary knots in the Helix nebula are uncertain. Are they created in the AGB wind, in which case they would be described as primordial, or in the planetary nebula, i.e. after the AGB phase? A primordial nature of the globules was suggested by \citet{dyson1989} where relatively massive density enhancements are created in the atmosphere of the AGB star and somehow survive to the PN transition. To test this theory, \citet{huggins2002small} looked for the presence of globule-like structures in the molecular halo of the young planetary nebula NGC 7027 and in the atmosphere of the AGB star IRC+ 10216  without success. On the other hand, non-primordial globules would imply that the physical conditions inside the nebula fostered the creation of cool molecular clumps. \citet{matsuura2009} analysed the likelihood of the two scenarios and reported that part of the H$_2$ in the knots could be primordial, but the formation and survival of H$_2$ in dense and ionized regions inside PNe are also possible, as investigated by \citet{aleman2004}. The effect of spiral wave compression from the interaction with a binary companion could also result in the formation of clumps in the AGB wind.
\subsubsection*{Tail formation}
\begin{figure*}
	\includegraphics[width=\textwidth]{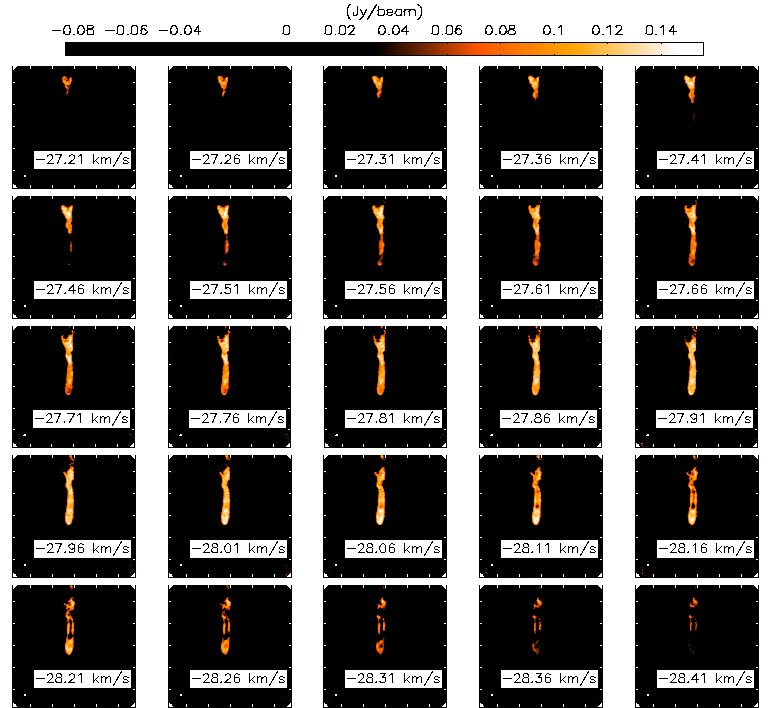}
    \caption{$^{12}$CO(2--1) channel maps of the CO globule centred at the velocities in the bottom left of each channel. The channels are $0.05\,\textmd{km}\,\textmd{s}^{-1}$ wide. y and x axes are the same as in Fig. \ref{fig:integrated_CO}.}
    \label{fig:velocityCO}
\end{figure*}
\begin{figure*}
	\includegraphics[width=\textwidth]{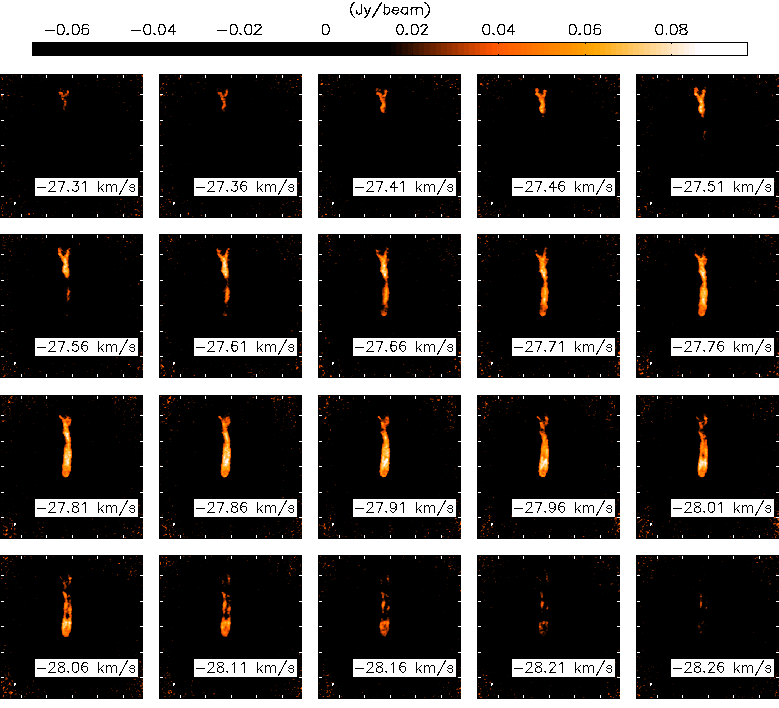}
    \caption{Same as Fig. \ref{fig:velocityCO} for $^{13}$CO(2--1).}
    \label{fig:velocity13CO}
\end{figure*}
\begin{figure*}
\begin{tabular}{ll}
\small \textbf{(a)}
\includegraphics[scale=0.4]{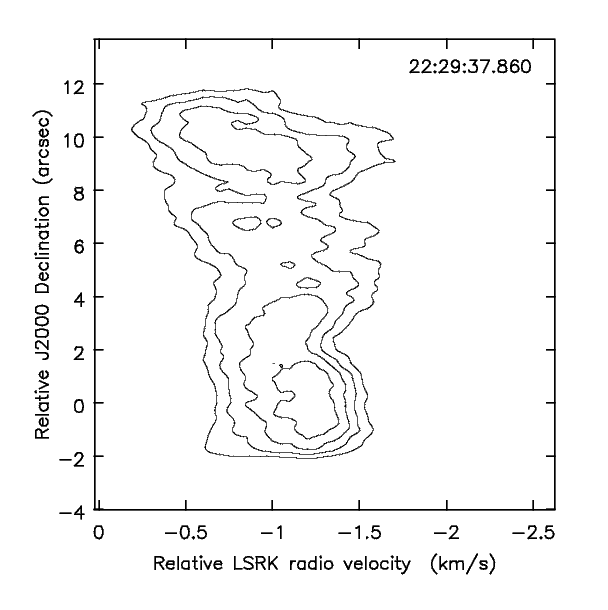}
&
\small \textbf{(b)}
\includegraphics[scale=0.405]{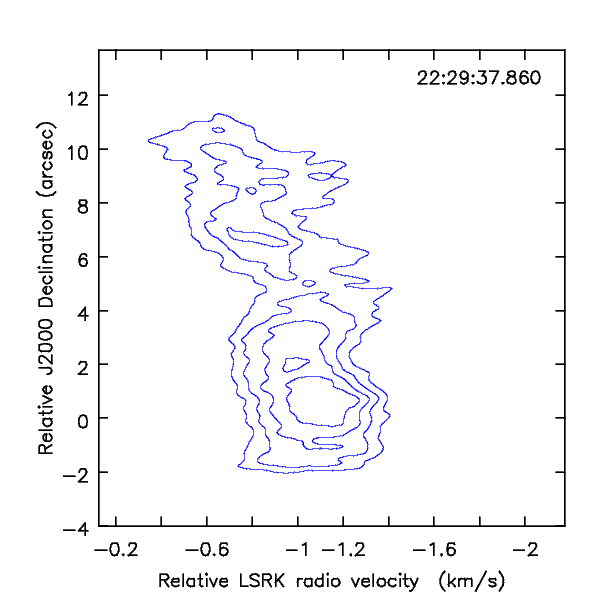}

\end{tabular}
\caption{Right ascension averaged velocity strip maps of the C1 globule along the declination axis and centred at right ascension given in the top right of each map. \textbf{(a)}:  $^{12}$CO(2--1) and contour intervals are $0.0108\,\textmd{Jy}\,\textmd{beam}^{-1}$. \textbf{(b)}: $^{13}$CO (2--1) and contour intervals are $0.006\,\textmd{Jy}\,\textmd{beam}^{-1}$.}
\label{fig:velocity_dec_CO}
\end{figure*}
\begin{figure*}
\centering
\begin{tabular}{c @{\qquad} c }
\small \textbf{(a)}
\includegraphics[scale=0.42]{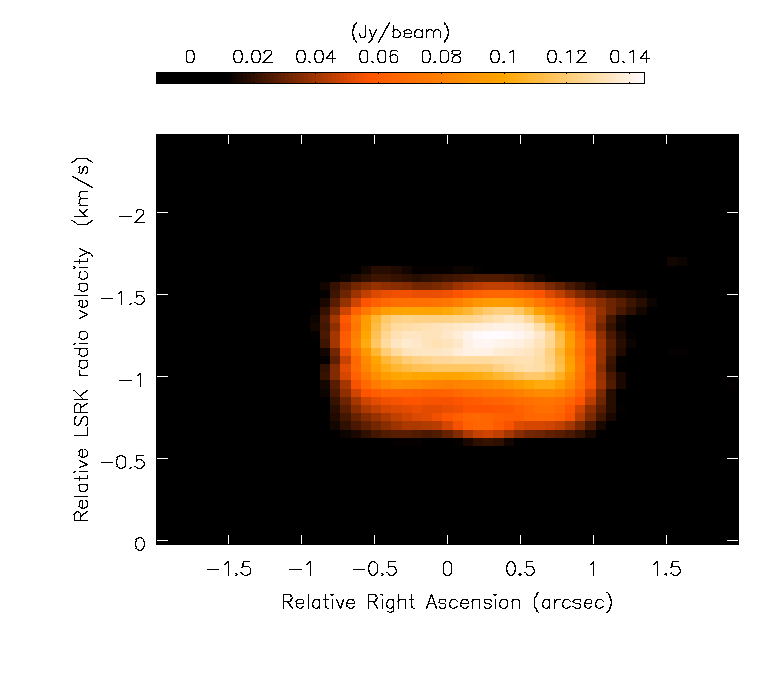}
&
\small \textbf{(b)}
\includegraphics[scale=0.42]{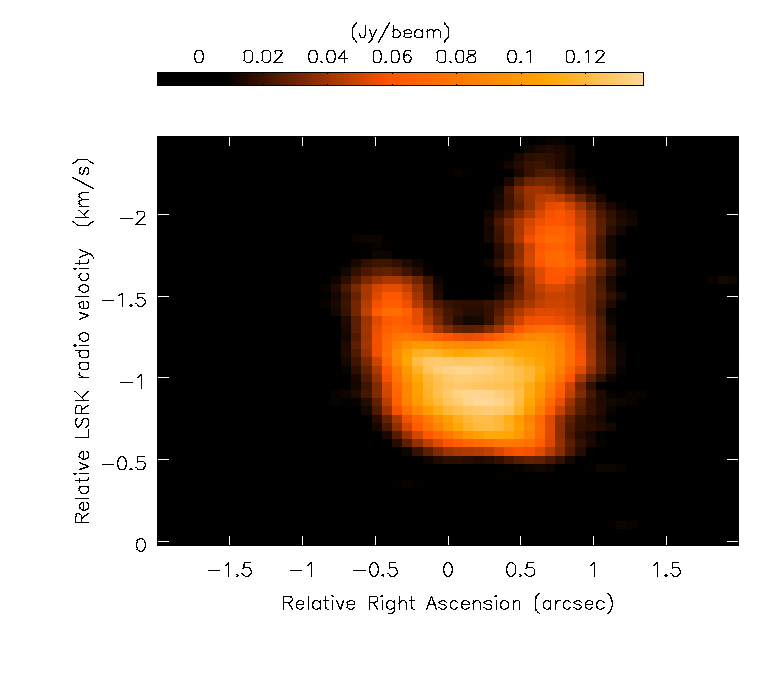}
\\
\small \textbf{(c)}
\includegraphics[scale=0.42]{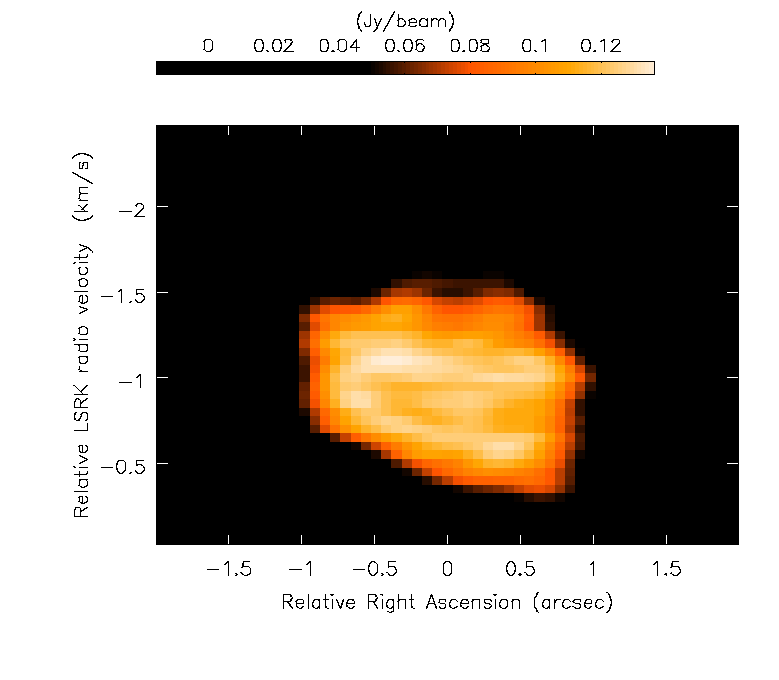}
&
\small \textbf{(d)}
\includegraphics[scale=0.42]{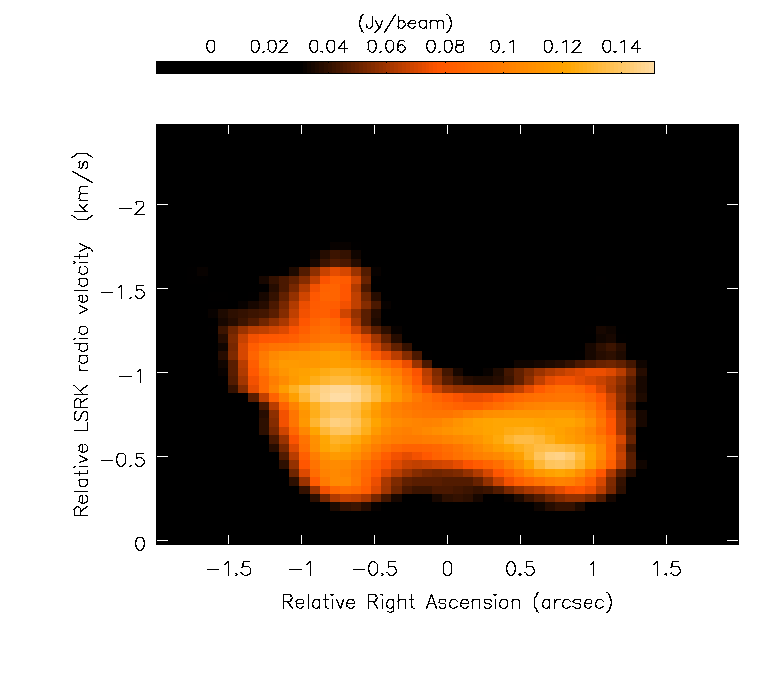}
\end{tabular}
\caption{Velocity against right ascension at different declinations for $^{12}$CO. \textbf{(a)}: head of the globule at a declination of $-0.5^{\prime\prime}$ (see Fig. \ref{fig:integrated_CO} for position reference), \textbf{(b)}: mid-tail at $+\,5.5^{\prime\prime}$, \textbf{(c)}: tail at $+\,9^{\prime\prime}$ and \textbf{(d)}: extended tail at $+\,10.5^{\prime\prime}$.  }
\label{fig:Vel_ra_cut}
\end{figure*}

Different models have been proposed to describe the formation of globule tails, the most common ones being the stream, the photoevaporation and the shadowing models. Those models are not intended to explain the origin of the density enhancement, i.e. whether they are primordial or not, but assume the presence of a dense core and describe the mechanisms that shape those density enhancements into comet-like objects. In the following, we will discuss those three models and see how consistent they are with our data.
\\
\\The stream model consists of an ambient wind and the flow from the knot interacting with each other. The velocity difference between the globule and the ambient wind results in a shock that creates the optical bow-shape of the head. The material from the head is swept by the wind to form the tail \citep{dyson2006}. The nature of the wind differs for different models. \citet{meaburn1996} suggested that expanding diffuse shells or the fast wind from the CS could interact with the dense cores and create the tails. \citet{meaburn1992} estimated the lifetime of globules and compared it to the lifetimes of the fast and expansion winds and found that both types of winds can potentially interact with the dense cores to produce the tails of cometary knots. Furthermore, an evolution in the nature of the ambient wind, from subsonic to supersonic, would lead to meandering in the tails of the globule. More recently, \citet{meaburn2010} described the stream model as particle winds from the CS sweeping past the slowly expanding system of knots. According to the stream model, the system of cometary knots has a typical radial expansion velocity of around $10\,\textmd{--}\,14\,\textmd{km}\,\textmd{s}^{-1}$ \citep{meaburn1996,meaburn1998}. From theoretical simulations of individual knots in [N$\,$\textsc{ii}], \citet{meaburn1998} suggested that the wind-swept model would produce an ablation flow velocity of $\sim10\,\textmd{km}\,\textmd{s}^{-1}$.
\\We look for a velocity gradient between the head and the tail of the C1 globule to test the ablation of the head by an ambient wind. Figure \ref{fig:velocityCO} and Fig. \ref{fig:velocity13CO} show the $^{12}$CO and $^{13}$CO(2--1) channel maps of C1 with a velocity width of $0.05\,\textmd{km}\,\textmd{s}^{-1}$.  In the $^{13}$CO channel maps, the tail is a few arcseconds shorter and observable in fewer channels than in $^{12}$CO. We notice that only the tail is visible in the maps at the top channels (at $\sim-27.2\,\textmd{km}\,\textmd{s}^{-1}$) which would suggest that there is head ablation. However, the head and the tail are both detected at velocities $\sim-28.3$ $\textmd{km}\,\textmd{s}^{-1}$. The centre of the tail has a velocity about $0.5\,\textmd{km}\,\textmd{s}^{-1}$ redshifted from the cocoon around it. Figure \ref{fig:velocity_dec_CO} illustrates the velocity distribution along the declination axis and shows a differential movement between the head and the tail of the globule, of around $0.4$ and $0.5\,\textmd{km}\,\textmd{s}^{-1}$ along the line of sight for $^{12}$CO and $^{13}$CO, respectively. This  velocity gradient is not enough to support a wind-swept pattern. Figure \ref{fig:Vel_ra_cut} shows the velocity along the right ascension axis for different declination for the $^{12}$CO emission. The velocity range is roughly constant in the head and the mid-region of the tail, but it presents a drop of $\sim 0.4\,\textmd{km}\,\textmd{s}^{-1}$ towards the extended tail, around $+\,10.5^{\prime\prime}$ from the tip of the head. Those values are far from the $10\,\textmd{km}\,\textmd{s}^{-1}$ ablation flow velocity expected and, thus, do not confirm the stream model. Our result is in agreement with \citet{huggins2002} who found an observable velocity gradient of $0.2\,\textmd{km}\,\textmd{s}^{-1}$ in the line of sight and of $0.8\,\textmd{km}\,\textmd{s}^{-1}$ in a radial direction. With a velocity flow of $\sim1\,\textmd{km}\,\textmd{s}^{-1}$, the end of the tail of the globule would be reached in around 10$^4\,$yr.

In addition, as mentioned in section \ref{sec:morphology} and \ref{sec:mass}, our results highlight the molecular nature of the tail of the C1 globule. Do the molecules in the tail originate from the head? Is the interaction strong enough to sweep the needed quantity of material from the head to create the heavily molecular tail? This model seems unlikely to explain our observations, in particular, the multipeak structure of the globule and the high mass of the tail. None the less, the meandering of the tail could be the consequence of streaming motion instabilities (e.g \citeauthor{dyson2006} \citeyear{dyson2006}). We conclude that the stream model alone does not describe the formation of the tail of the C1 globule.\\ 
\\From Fig. \ref{fig:ratio_12_13}, we observe that the highest $^{12}$CO/$^{13}$CO line ratio is in the extended tail (at $\sim+\,10^{\prime\prime}$). In section \ref{sec:origin}, we find that the velocity in this region is not the same as the velocity observed in the rest of the globule, which may indicate that the corresponding radiative environments are different. Therefore, we do not exclude the possibility of the observed extended tail to represent a different globule.
\\
\\\citet{canto1998} proposed the shadowing by density enhancement model to explain the origin of the cometary tails of the knots in the Helix nebula. In this model, the initial dense core can be large and is located in the photoionized region of the nebular envelope. The UV radiation from the CS hits the head but does not directly reach the tail region which is shadowed. The tail of the core is then illuminated by the diffuse radiation. In this model, the shadow region has a neutral inner core with a bright rimmed trailing envelope. 
\\Two versions of the globule shaping process by shadowing are possible \citep{huggins1992}. The first one consists of the shadowing of preexisting molecules close to the central star overrun by the ionization front of the nebula leading to the creation of molecular tails. The second version assumes that the tail and the head are created simultaneously. The latter would be more consistent with our data  given the high molecular mass of the tail.
\\The shadowing model can explain the molecular nature of the tail that we observe in CO, and could also explain the presence of the different peaks in the globule. Moreover, the flux of the radiation from the CS of the Helix nebula \citep{kaler1990}  is low enough to permit the neutral core to be shielded \citep{huggins1992}. 
\\
\\In the photoevaporation model \citep{lopez2001}, the UV photons from the central star interact with the material of the density enhancement. The surface of the core is heated by the UV radiation and the shock from that interaction creates the crescent tip of the head of the knot observed in optical images (see Fig. \ref{fig:overlay_HST}). The globule slowly evaporates as the nebula expands \citep{huggins1992}. However, this model does not clearly explain the formation of the tail. It is often combined with other processes such as the stream model, where a wind sweeps the photoionized gas and the neutral material to the tail (\citeauthor{huggins1992} \citeyear{huggins1992}, \citeauthor{matsuura2009} \citeyear{matsuura2009}).
The photoevaporation model alone does not explain the molecular properties of the knot.
\\
\\The peaks observed in the body of the globule can also be due to the interaction of globules with one another \citep{pittard2005}.  In the case of the C1 globule, it seems to be fairly isolated. However, the peak  observed at $\sim$+$\,8^{\prime\prime}$ (see Fig. \ref{fig:integrated_CO}) can be the relic of the condensation of an other globule. 
\\
\\ \citet{matsuura2009} investigated a certain number of knots in molecular  H$_2$(1-0) and concluded that the stream and photoevaporation models have important contributions to the tail shaping process at the expense of shadowing, whose level of contribution had yet to be proven. However, the shadowing model is the one that best explains our data, as it is the only model that is consistent with a highly molecular tail. A tail formation scenario where the effects of all models are taken into account is desirable, so, we suggest the following combination of models. The initial density enhancement is a large dense core made of gas and dust whose head is hit by UV radiation from the CS, leading to the photoevaporation and shaping of the ionized bow-like head, while the remaining part of the knot is shielded. The diffuse radiation from the head illuminates the shadowed tail and the globule slowly evaporates. This is consistent with Fig. \ref{fig:overlay_HST}, showing an \textit{HST} optical observation of the ionized head overlaid with our CO data of the C1 globule. Though we did not find any evidence of ablation of the head to support the stream model, we do not exclude the possibility of it being responsible for the meandering of the tail. 
\subsubsection*{Near and distant knots}
Globules near and far from the CS present different physical appearances. Globules closer to the CS, like the C1 globule, exhibit a head and a defined extended tail, whether in molecular (H$_2$) or in [O$\,$\textsc{iii}] (in absorption), whereas those in the far side do not present such defined tails and are more numerous \citep{matsuura2009}. In addition, the bow shape of the ionized head is not always observed for knots located far from the CS. The question of which of those categories represent the younger globules, i.e. their evolution, is directly related to the physical processes responsible for their creation and shaping. Our analysis shows that shadowing and photoevaporation are the main contributors to the shaping of the C1 globule. This may explain why the knots in the outer part do not present visible tails in optical images ( e.g. \citeauthor{o2004} \citeyear{o2004}), and lie in dense regions where tails are blended in and indistinguishable in H$_2$, as imaged by \citet{matsuura2009}. In fact, the high number density of knots in the outer region is expected to make the shielding effect more efficient and, combined with interactions due to the proximity of the globules with one another, reduce the possibility of observing distinct tails. However, their morphological differences have also been suggested to indicate that different processes may be responsible for the shaping of the globules in the inner and outer parts of the nebula \citep{matsuura2009}.
\section{Conclusion}
\label{sec:Conclusion}
We derived the physical parameters of the C1 globule from new CO observations. With an optically thick $^{12}$CO line, we found that the molecular mass of the knot may have been highly underestimated by previous studies. The globule has a molecular mass of  M$_{\textmd{knot}}$=$1.8\,\times10^{-4}\,$\(\textup{M}_\odot\), which is mainly carried by the tail. With such a mass, we suspect the nebula to be, in fact, condensation of globules, where the knots close to the CS such as C1 are more massive than the ones in the outer region. From dust observation, we find a gas mass of $5.5\times10^{-6}\,$\(\textup{M}_\odot\).
The optical depth-corrected $^{12}$C/$^{13}$C and $^{16}$O/$^{18}$O ratios are $\sim$10 and $\sim$115, respectively. Those values are not consistent with the carbon and oxygen isotopic ratios of AGB carbon stars. The $^{12}$CO optical depth may be underestimated. The high derived molecular mass and the low CO isotopic ratios could be the effects of clumping in the globule. Investigating the likelihood of the presence of clumps in C1 would require higher resolution data. The tail formation process is also discussed and our findings are in agreement with the shadowing model combined with photoevaporation. Though our observations did not show strong evidence to support the stream model, we do not exclude its possible contribution to tails meandering. Our data therefore suggest that the shaping of the knots into comet-like structures is likely to happen after photoionization of the nebula, i.e. not during the AGB phase. Our model seems to be compatible with ionized gas observations of the C1 globule, and could also explain why cometary knots near and distant from the CS appear to have different morphologies. Finally, given the multi-peak structure of the globule, the feature observed in the optical depth maps, the high mass of the tail and its slightly shifted velocity, it is possible that the C1 globule actually consists of two different knots. 

\section*{Acknowledgements}
The authors would like to dedicate the paper to Patrick Huggins. He obtained the ALMA observations as PI, but sadly died before seeing the data.  The data has previously been studied by Wouter Vlemmings and Mikako Matsuura, and their unpublished work affected this research.\\
A.M. is funded by the UK's Science and Technology Facilities Council (STFC) through the DARA (Development in Africa with Radio Astronomy) project, grant number ST/M007693/1. A.A. is funded by the STFC at the UK ARC Node.  A.A.Z acknowledges funding by the STFC under grant  ST/P000649/1.\\
This paper makes use of the following ALMA data: ADS/JAO.ALMA\#2012.1.00116.S. ALMA is a partnership of ESO (representing its member states), NSF (USA) and NINS (Japan), together with NRC (Canada), MOST and ASIAA (Taiwan), and KASI (Republic of Korea), in cooperation with the Republic of Chile. The Joint ALMA Observatory is operated by ESO, AUI/NRAO and NAOJ. 
\\This work has made use of data from the European Space Agency (ESA) mission
{\it Gaia} (\url{https://www.cosmos.esa.int/gaia}), processed by the {\it Gaia}
Data Processing and Analysis Consortium (DPAC,
\url{https://www.cosmos.esa.int/web/gaia/dpac/consortium}). Funding for the DPAC
has been provided by national institutions, in particular the institutions
participating in the {\it Gaia} Multilateral Agreement.




\bibliographystyle{mnras}
\bibliography{CO_in_the_C1_globule.bib} 





\bsp	
\label{lastpage}
\end{document}